# A Floating Octave Bandwidth Cone-Disc Antenna for Detection of Cosmic Dawn

[1]Agaram Raghunathan, [2]Ravi Subrahmanyan, [1]N. Udaya Shankar, [3]Saurabh Singh, [4]Jishnu Nambissan,
[1]K.Kavitha, [5]Nivedita Mahesh, [1]R. Somashekar, [1]Gaddam Sindhu, [1]B. S. Girish, [1]K. S. Srivani, [1]Mayuri S. Rao

[1]The authors are with the Raman Research Institute, Sadashivanagar, Bangalore 560080, India (e-mail: raghu@rri.res.in; uday@rri.res.in; kavitha@rri.res.in; som@rri.res.in; sindhu@rri.res.in;bsgiri@rri.res.in; vani_4s@rri.res.in; mayuris@rri.res.in).
[2]Ravi Subrahmanyan was with the Raman Research Institute, Sadashivanagar, Bangalore-560080, India. He is now with CSIRO Astronomy & Space Science, PO Box 1130, Bentley WA 6102, Australia (e-mail: ravi.subrahmanyan@csiro.au).
[3]Saurabh Singh is with the McGill Space Institute, McGill University, 3550 rue University, Montreal, QC H3A 2A7, Canada, on leave from the Raman Research Institute, Sadashivanagar, Bangalore 560080, India (e-mail: saurabhs@rri.res.in).
[4]Jishnu Nambissan is with the Curtin Institute of Radio Astronomy, Turner Ave, Bentley WA 6102, Australia, on leave from the Raman Research Institute, Sadashivanagar, Bangalore 560080, India (e-mail: jishnu@rri.res.in).
[5]Nivedita Mahesh is with the School of Earth and Space Exploration, Arizona State University, Tempe, Arizona 85287, USA (e-mail: nivedita.mahesh@asu.edu).

*Abstract*—The critical component of radio astronomy radiometers built to detect redshifted 21-cm signals from Cosmic Dawn is the antenna element. We describe the design and performance of an octave bandwidth cone-disc antenna built to detect this signal in the band 40–90 MHz. The Cosmic Dawn signal is predicted to be a wideband spectral feature orders of magnitude weaker than sky and ground radio brightness. Thus, the engineering challenge is to design an antenna at low frequencies that is able to provide with high fidelity the faint cosmological signal, along with foreground sky, to the receiver. The antenna characteristics must not compromise detection by imprinting any confusing spectral features on the celestial radiation, ground emission or receiver noise. An innovation in the present design is making the antenna electrically smaller than half wavelength and operating it on the surface of a sufficiently large water body. The homogeneous and high permittivity medium beneath the small cone-disc antenna results in an achromatic beam pattern, high radiation efficiency and minimum unwanted confusing spectral features. The antenna design was optimized in WIPL-D and FEKO. A prototype was constructed and deployed on a lake to validate its performance with field measurements.

*Index Terms*—Antenna measurements, radio astronomy, reflector antennas.

## I. Introduction

Modern cosmology has, over the last few decades, made considerable progress in building a detailed model for the dynamical evolution of the Universe. The key observational constraint which refined theoretical model was cm and mm wavelength measurements of the cosmic microwave back- ground. However, astrophysical aspects of the evolution like formation of the first stars and Galaxies is largely unconstrained [1]. A key method for resolving this problem is detection of the redshifted 21-cm line from neutral hydrogen

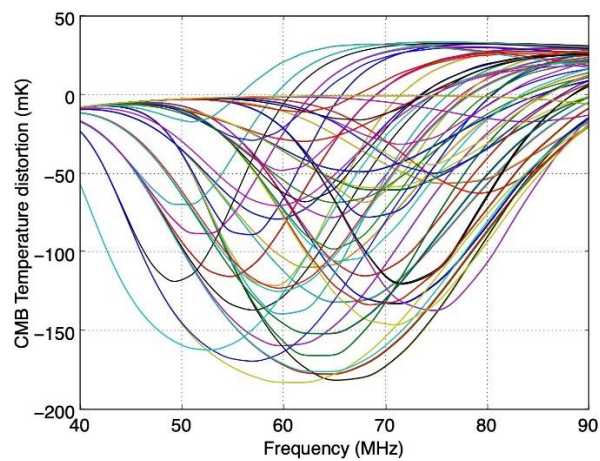

Fig. 1. Examples of signals from Cosmic Dawn drawn from the atlas of templates in Cohen et al. [3]. These absorption dips are predicted in the cosmic microwave background in the frequency range 40–90 MHz. Different dip profiles represent different plausible scenarios of formation of first stars; all these scenarios are currently allowed by observational constraints.

atoms in that uncertain epoch, often called Cosmic Dawn [2]. This is a global signature appearing as an absorption dip in the spectrum of the cosmic microwave background in the frequency range 40–90 MHz. All the parameters of the profile like strength, location are related to the physics at Cosmic Dawn. Examples of signal profiles allowed by current theory are shown in

Figure 1; the diversity in the predictions is indicative of weak constraints on astrophysical parameters by available observations [3].

Several radiometer designs have been trialed and observations made at radio quiet sites, yielding useful constraints [4][5][6][7], and there is also a claim of detection of an absorption profile from cosmic dawn by the EDGES low- band system [8]. However, the absorption depth inferred is greater than the maximum allowed in standard cosmology and has been questioned [9][10][11]; therefore, the claim awaits confirmation from independent observations. The detection of the faint cosmic dawn signal has been difficult and results so far have been contentious due to the difficulty in designing adequately sensitive radiometers that either avoid or calibrate out systematics. The critical component of precision radiometers is the antenna element that couples the real world to the laboratory-tested receiver.

The EDGES radiometer [12] used an adaptation of the four-point antenna [13][14], which was a planar dipole over a ground plane and covered the 100–200 MHz band. More recently, EDGES used scaled versions of wideband blade antenna [8] for radiometer measurements in the 50–100 and 90–200 MHz bands. Drooped down crossed dipoles operating in the 30–85 MHz band were used in LEDA radiometers [15][16]. Cross dipoles in the form of the HIbiscus antenna [17] was used as the electromagnetic sensor for the SCI- HI radiometer [18] and also the PRIzM radiometer [19]. Frequency independent 2-arm conical spiral antennas were adopted for the radiometers of CoRE [20] and also BigHorns [21].

The SARAS radiometer [22] used a wideband fat-dipole [23] over ferrite absorbing ground screen, and operated over the band 87.5–175 MHz. That was replaced in SARAS 2 [4][5] by a spherical monopole over ground [24] to operate over 110–200 MHz. In this manuscript, we describe a new purpose design of a Cone-Disc antenna to operate in the frequency band 40–90 MHz and serve as the antenna for the SARAS 3 radiometer.

The structural parameters of the Cone-Disc antenna were modeled using WIPL-D and FEKO to achieve monotonic and extremely smooth characteristic in both transmission and gain versus frequency. The antenna was designed for operation on an electromagnetically transparent raft floating on the surface of a lake. This design choice of having water as the medium immediately beneath the antenna to considerable depth was motivated by the reasoning that (i) water potentially provides an extremely homogeneous medium below, unlike real ground that may have structure and layering beneath that may influence performance, (ii) water as a dielectric beneath the antenna can be accurately modeled, and (iii) the antenna over water— instead of ground—would achieve higher radiation efficiency because of the higher electrical permittivity of water relative to soil.

The antenna was constructed and its parameters were refined in field tests where it was deployed on a raft on a lake. Section II describes the design objectives and philosophy. Design details are provided in Section III and a discussion on the effects of floating the antenna on water is in Section IV. Section V describes the fabrication of the antenna. Performance characteristics are given in Section VI and a summary is in Section VII.

## II.  DESIGN OBJECTIVES AND PHILOSOPHY

The objective of our antenna design is to detect spectral features imprinted in the cosmic microwave background in the frequency range 40–90 MHz. These features are predicted to be $10^5$ times fainter than radio emissions of our Milky Way and discrete radio sources in the sky. The celestial radio emission, in particular from the Milky Way, varies significantly in brightness temperature across the sky and has substantial spatial structure on a variety of scales. The structure in the Galactic and extragalactic radio sky requires that any antenna aiming to detect Cosmic Dawn has frequency independent characteristics. Beam chromaticity translates spatial structure in sky emission to spectral structure, which may confuse detection of the faint Cosmic Dawn signal. This places stringent limits on tolerable chromaticity in the beam.

A second critical characteristic of a cosmic dawn antenna is the S11 scattering matrix element. The sky signal propagating through the antenna is shaped by this scattering matrix element. Additionally, the low-noise amplifier behind the antenna has noise waves propagating in both forward and reverse directions, and the reverse propagating wave is reflected and shaped by the antenna S11. Therefore, any complexity in the spectral shape of S11 makes the multiplicative spectral gain and additive receiver noise to have complex band shapes. As a consequence, the Galactic and extragalactic foregrounds and receiver noise may be inseparable from any faint cosmic dawn signal, thus compromising detection.

Other potential sources of errors in radiometer measurements of Cosmic Dawn are the antenna radiation efficiency, ground emission that couples into the antenna, antenna resistive losses and balun losses, all of which may have complex spectral characteristics that may require calibration methods or else design that can marginalize their effects.

In many ongoing experiments such as EDGES [8][12], LEDA [15], SCI-HI [18] and PRIzM [19], even though the dipole antennas are designed to be frequency independent to a large extent, their residual frequency dependencies in both S11 and beam chromaticity cannot be marginalized in the data analysis. Hence these experiments introduce unavoidable calibration steps which constitute additional sources of errors in the detection sensitivity.

Conical spiral antennas with large structural bandwidth were adopted by the CoRE [20] and BigHorns [21] experiments. Although their beam patterns are achromatic, their large electrical lengths resulted in complex structures in S11 which are difficult to calibrate. It is, of course, in principle possible to characterize these potential errors. However, the SARAS approach has been to purpose design the antenna to keep these errors, arising from chromaticity and complex structure in S11, sufficiently small and thus avoid these calibrations.

The SARAS 1 fat dipole [23] departed from other dipole designs in that it was less than half wave long at the highest frequency of the operating band, and used ferrite absorber tiles as a ground plane. Thus the beam was achromatic. However the performance was limited by frequency dependent characteristics of the balun, absorptivity of ferrite tiles and standing waves between the antenna and a following low-noise amplifier.

SARAS 2 moved from dipoles to spherical monopoles, thus bringing the antenna terminals to ground level and at the amplifier inputs [24]. The improved design resulted in (i) achromatic beam patterns and (ii) made the antenna free from any balun and associated lossy elements. However, it shape of the antenna $S11$ was complex due to interaction with had a significantly reduced efficiency and, importantly, the inhomogeneous ground.

SARAS 3 improves further the design of the monopole by changing to a conical element for the monopole and also avoiding many of the limitations arising from the coupling to soil by floating the antenna on a large homogeneous body of water. The design presented here is optimized for the 40– 90 MHz band. As was the case for SARAS 2, the height of the conical monopole of SARAS 3 as well as the radius of the metallic ground plane are made shorter than half a wavelength at the highest operating frequency, thus providing beams with low chromaticity. Second, the conical shaped monopole element along with floating the antenna on a substrate with high dielectric constant improves efficiency.

Conventional wideband designs allow both return loss and gain to vary in a complex manner and thus achieve optimal performance over a wide band, which results in non-smooth spectral response for both antenna gain and return loss [25]. The considerations and priorities in our antenna design are different from that driving conventional wideband designs. For example, it is more important to have ultra-smooth return loss characteristic, even if the return loss is not substantial across the wide band. The reason why smoothness in return loss is vital in the case of antenna designs for cosmic dawn is that complexity in this characteristic could potentially limit sensitivity through systematic errors, since calibration is a challenge at the precision needed for the science goal. On the other hand, loss in efficiency could be made up for by increased observing time.

### III. DESIGN DETAILS

The design of the SARAS 3 antenna followed several guiding principles to meet the science requirements and have good impedance match and smooth spectral response over a wide bandwidth. The radiating structure was made less than half wavelength at the highest operating frequency so that the antenna is endowed with frequency independent characteristics. The structural dimensions were made to vary uniformly from the feeding point in order to minimize surface current reflections at structural discontinuities, which have the potential to result in undesirable features in the spectral response. The environment was carefully considered and included in the design and modeling, since any conducting element and substrates would influence the antenna performance. The antenna structure was made rugged to retain its shape and achieve consistent performance over long-duration observations in the field. The antenna was designed to have minimum gain towards horizon so that it is less sensitive to terrestrial man-made radio frequency interference. Finally, and most importantly, the design aims for both antenna impedance and gain to have smooth characteristic across the band, so that they may be modeled with functional forms and marginalized without compromising detection of the Cosmic Dawn signal. Within the family of monopole antennas over a finite ground plane, different types of low profile monopole structures, like spheres, (ii) inverted cones, and (iii) profiled monopoles, were investigated for their electrical characteristics. Among them, the structure consisting of an inverted cone over a circular disc was preferred. This structure has a uniformly varying dimension and consequently displays characteristics that have relatively smoother change with frequency; hence systematics and calibration complexities are inherently reduced. As dis- cussed above, this aspect is of greater importance in selection of the antenna structure compared to total efficiency, which may be better in structures that are appropriately profiled. However, the downside to substantial reduction in antenna efficiency is a corresponding reduction in tolerance to internal spurious and hence the design of the analog receiver and digital spectrometer become more challenging.

A schematic of a cone-disc antenna is shown in Figure 2. It has three primary structural parameters: (i) reflector radius,
(i) slant height and (iii) semi-cone angle, which control the overall performance of the antenna. The structure at the feed point is described by (iv) radius of the cone base and (v) size of the gap between reflector and cone base.

**Reflector radius**: Radial surface currents on the reflector plate meet an impedance discontinuity at the outer edge of the disc, and are partially reflected back to the feed point. This results in the scattering matrix element $S11$ spectral characteristics, with period $c/(2R)$, where $R$ and also the reflection efficiency to have a ripple in their is the radius of the reflector. To avoid complexity in these antenna characteristics we choose to design the conductive reflector to be small enough so that the octave operating band contains less than a quarter cycle of ripple.

**Slant height**: As in the case of the reflector plate, so also the currents on the surface of the cone reflect at the top edge and flow back towards the feed point resulting in spectral ripples in characteristics. The slant height is also to be kept sufficiently small so that at most a quarter of a ripple period appears in the band. We select the slant height of the cone to be the same as the radius of the reflector plate.

**Semi-cone angle**: This parameter has a significant impact on the antenna impedance, and hence its match to the low-noise amplifier connected at its terminals, radiation pattern and their variation with frequency over the wide operating band. This parameter is hence determined during the electromagnetic simulation analysis, to meet the design requirements discussed above.

Feed point structure: The radius of the base of the cone was made equal to the gap height between the reflector radius and base of the cone. Thus the surface of the cone, if extended to the reflector, will have its apex at the center of the reflector.

The electromagnetic (EM) modeling of the antenna in the frequency range of 40–90 MHz was carried out using electromagnetic modeling software tools WIPL-D and FEKO, to optimize its structural dimensions. Use of more than one soft-ware examines antenna characteristics with multiple tools thus providing a measure of the latitude in performance predictions arising from EM simulation algorithms and implementations. The SARAS 3 antenna was designed to be operated over the surface of water instead of soil because of its high and homogeneous dielectric constant, with $\varepsilon_r$ close to 80 (see Section IV for details of the effect of water below the cone-disc

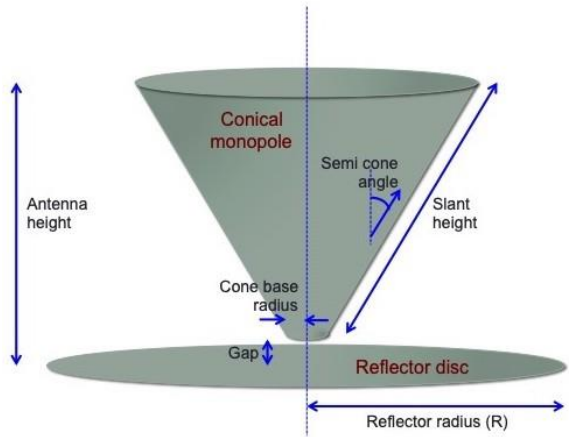

Fig. 2. Schematic giving the parameters that describe the structure of a cone-disc antenna.

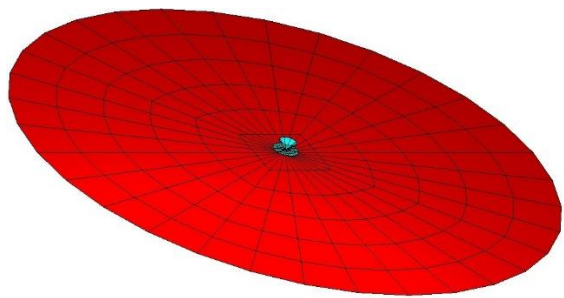

Fig. 3. Simulated structure of the Cone-disc antenna placed over the water surface. In blue is shown the antenna structure and in red is shown the surface of water extending out to 18 m.

antenna). This is expected to enhance the antenna performance and have smooth spectral response.

In the case of WIPL-D, the simulation box was extended to a radius of 18 m: this limitation was due to a limit in the number of unknowns that could be solved for in the implementation and was deemed sufficient since (i) the box size was an order of magnitude greater than the size of the reactive near field and (ii) because any erroneous boundary matching at that radius from the feed point would result in high order structure in the performance characteristics, well above that expected for the Cosmic Dawn signals. The water below the antenna was modeled as homogeneous and of infinite depth; effects of finite depth are considered below in Section IV. The relative dielectric constant $\varepsilon_r$ of the water was assumed to be 80 and conductivity $\sigma = 0.0022$ Siemens (S) m$^{-1}$. In the case of FEKO, water was modeled as a uniform half space boundary with water extending to infinity in depth and towards the horizon. First, the design frequency which determines the electrical dimensions of the cone-disc antenna was chosen to be above 100 MHz so that the antenna is less than half wavelength at the highest frequency in the operating band, which is fully below 100 MHz. This also ensures that within the operating band the characteristics would be smooth without any convoluted spectral structure arising due to any in-band resonance.

Optimization using EM simulation was carried out by adopting the 'variation of parameters' technique in which every parameter was varied, one at a time, to understand its effect on the performance of the antenna. The primary goal was spectral smoothness in return loss characteristics and frequency independent beam patterns, with high efficiency throughout the band as a secondary goal. A given spectral response is said to be smooth when it does not leave residuals that might confuse the signal being detected, when fitted using a low order polynomial. The parameter space for the antenna structure was explored around a set of nominal parameter values, which are given in the first column of Table I. The model simulated is shown in Figure 3. The receiver at the antenna base is assumed to be housed within a square aluminum box, of side 0.52 m and height 0.14 m, located beneath the circular disc at its center; the box is included in EM model. The antenna is assumed to be placed on an electromagnetically transparent raft with the reflector plate at a fixed height above the surface of water, which is nominally 200 mm. Initially optimization was carried out by varying the design frequency over a limited range: 130, 150 and 170 MHz. The reflection efficiencies ($1 - |S_{11}|^2$) obtained for various design frequencies are shown in Figure 4. We next compute the expected distribution in antenna temperature across the operating band of the antenna. For this, we adopt the reflection efficiencies shown in Figure 4, radiation efficiency and antenna beam pattern as discussed below, and adopt GMOSS [26] as a model for the sky spectral brightness. The average antenna temperature that is expected to be measured by the antenna, for the various design frequencies, is shown in Figure 5.

TABLE I
PARAMETERS DEFINING THE CONE-DISC ANTENNA DESIGN.

| Structure Parameter | Nominal Value | Adopted Value |
|---|---|---|
| Design Frequency | | 150 MHz |
| Reflector radius | 0.8 m | 0.83 m |
| Slant height | 0.8 m | 0.83 m |
| Semi-cone angle | 45° | 45° |
| Gap height at base of cone | 1 mm | 1 mm |
| Cone base radius | 1 mm | 1 mm |
| Height of reflector from water surface | 0.2 m | 0.2 m |

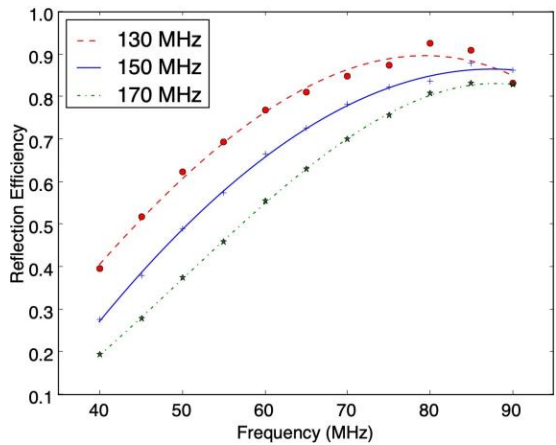

Fig. 4. Reflection Efficiency versus frequency for different choices of design frequency. In this figure as well as in following figures the symbols are the simulation results and the continuous lines are low-order polynomial fits that smooth over the errors arising from the finite size of the simulation box.

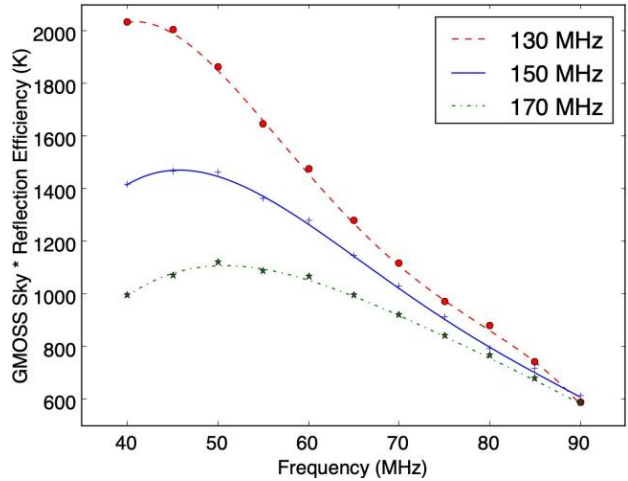

Fig. 5. Expected antenna temperature versus frequency for different choices of design frequency, assuming a model sky brightness distribution given by GMOSS [25].

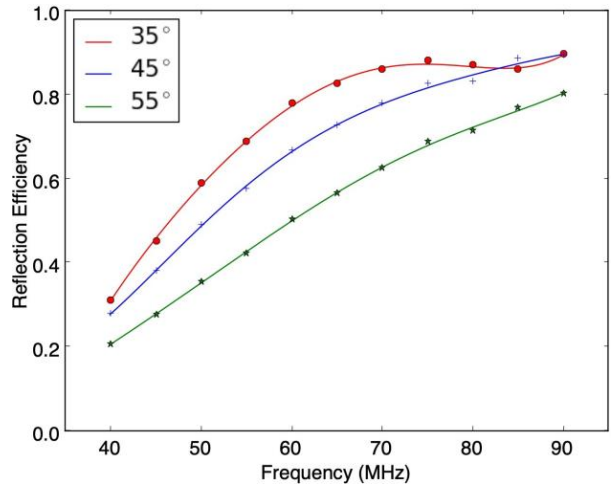

Fig. 6. Effect of varying the semi-cone angle on the reflection efficiency characteristics of the cone-disc antenna.

Selection of the design frequency was aimed at (i) making the spectral response of the antenna free from any embedded ripples and (ii) ensuring higher reflection efficiency. A design frequency of 130 MHz brings the resonance of the antenna within the band. Therefore, moving the design frequency to 150 MHz or beyond is necessary to keep the resonance out of the band. However, higher design frequencies drop the reflection efficiency, which is also undesirable. Additionally, design frequencies well above 150 MHz result in an inflection in the efficiency curve within the band as the efficiency begins to have an upturn at the lower frequencies within the band.

If we assume that the antenna would be used over the octave band 43.75–87.50 MHz, thus avoiding the band allocated to FM, then the bandwidth would be 43.75 MHz. As discussed in Section III above, the slant height of the antenna was kept the same as the radius of the reflector. During the simulation, these parameters ( reflector radius and slant height) were optimized to be equal to 0.415 times the wavelength at the design frequency. For design frequencies 130 MHz, 150 MHz and 170 MHz, the above physical parameters were 0.95 m, 0.83 m and 0.73 m respectively. To avoid accommodating more than quarter cycle of ripple in the band from back reflections of receiver noise at the outer edge of the reflector plate, the reflector radius needs to be limited to 0.86 m. This excluded design frequencies below about 150 MHz. In view of these considerations, the design frequency was adopted to be 150 MHz.

With the optimized values of reflector radius and slant height, the semi-cone angle of the antenna was varied. Reflection efficiency versus frequency for semi-cone angles of 35°, 45° and 55° are shown in Figure 6. The simulation results indicate that the antenna with smaller semi-cone angle will have greater efficiency. However, smaller cone angles cause (i) the resonance of the antenna to move down into the band and (ii) the functional form of the antenna transfer function will be of higher order curvature. Therefore, smaller cone angles potentially require higher order functional forms for description and are more likely to confuse detection of faint Cosmic Dawn signals embedded in the sky spectrum. Larger cone angles reduce the reflection efficiency across the band. On balance, the cone-disc antenna was given a semi-cone angle of 45°.

In Figure 7 we show the effect of changing the height of the antenna above the surface of water. If the height of the circular disc above the surface of water is less than 200 mm, the resonance moves into the band. Antennas with heights more than 200 mm would be expected to have smooth reflection efficiencies; however, they would have lower efficiencies throughout the band. Thus we select the height to be 200 mm. The selected values of the structural parameters for the SARAS 3 cone-disc antenna are in the second column of values in Table I. The gap and the radius of the cone base were each set at 1 mm. the antenna—the amplitude of the $S11$ scattering matrix. The reflection coefficient characteristics expected of element is shown in Figure 8. There is some deviation in the expectations from FEKO compared to WIPL-D; however, they are qualitatively the same and this aspect of spectral smoothness is critical for the science goal. The differences observed in the simulation results of WIPL-D and FEKO are attributed to the approaches of each one of them in their computation. FEKO is based on triangular mesh approach where as WIPL-D uses a quad mesh. FEKO uses a lower order MoM compared to WIPL-D. FEKO assumes infinite ground during simulation; however, WIPL-D can only work with finite dimensions for ground and, in this particular case, water. Accuracies of both FEKO and WIPL-D are different.

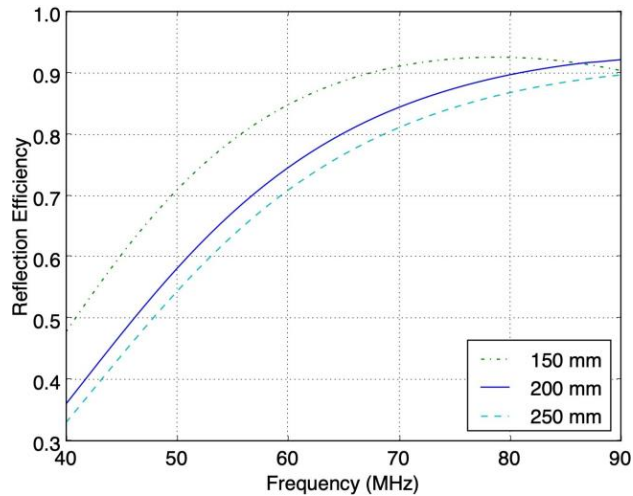

Fig. 7. Effect of varying the height of the antenna disc above the water surface, on the reflection efficiency characteristics of the cone-disc antenna.

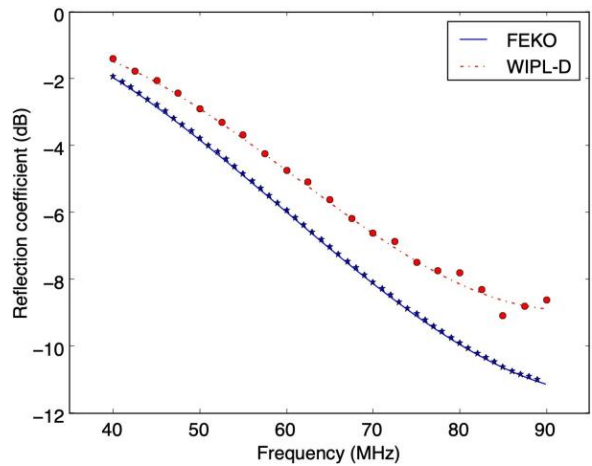

Fig. 8. Reflection coefficient expected for the SARAS 3 antenna from WIPL-D and FEKO modeling.

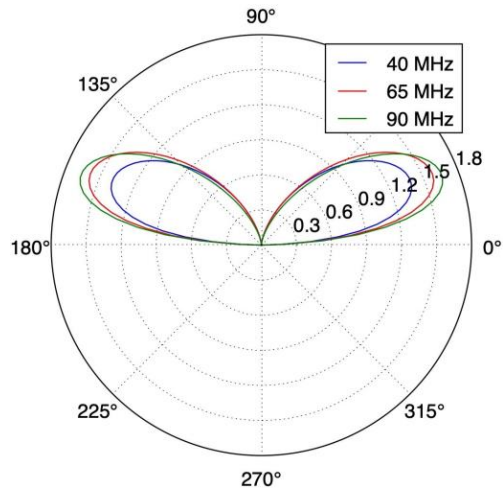

Fig. 9. Radiation patterns of the SARAS 3 antenna at 40, 65 and 90 MHz, from the WIPL-D EM modeling.

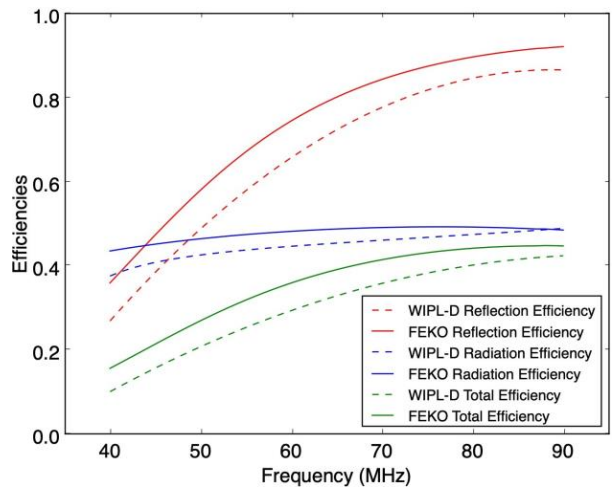

Fig. 10. Reflection, Radiation and Total efficiency expected for the SARAS 3 antenna from WIPL-D and FEKO modeling

The expected radiation patterns for the SARAS 3 cone-disc antenna are shown in Figure 9. The beams are omni-directional. Since the monopole height is less than quarter wavelength at the highest operating frequency, the patterns are similar at different frequencies and have no side lobes; there are nulls towards zenith and towards horizon. The variation in gain over the frequency band has been corrected for the reflection efficiency and hence represents the change in radiation efficiency, which increases with frequency. The beam is slightly chromatic owing to the finite size of the ground plane: the peak of the beam shifts somewhat towards lower elevation with increasing frequency. At 40 MHz the peak is at an elevation angle of $23.8°$, $22.8°$ at 65 MHz and $21.5°$ at 90 MHz, thus moving towards horizon almost linearly by about $2.3°$ across the 40–90 MHz band. This behavior of the beam is the same in both WIPL-D and FEKO modeling. Simulations with GMOSS model sky have shown that this small level of chromaticity, together with the spatial structure in sky brightness distribution, does not lead to spectral structure that could confuse Cosmic Dawn signals.

The expected efficiencies of the antenna—reflection, radiation and total—are displayed together in Figure 10. The predictions of WIPL-D and FEKO modeling are shown. The predicted efficiencies are somewhat lower in the case of WIPL-D; however, qualitatively the profiles are similarly smooth, which is the critical requirement of the purpose design.

## IV. Effect of floating the cone-disc antenna on a water body

The effect of water as a dielectric medium immediately below the antenna depends on its complex dielectric constant and conductivity. These determine the penetration depth for electromagnetic waves and hence the required vertical extent of water beneath the antenna, so that the impedance discontinuity at the boundary at the lake bed does not adversely affect performance.

Conductivity of sea water is about 5 S m$^{-1}$. Our sampling of water from inland lakes that are remote from townships yield conductivities in the range 0.002 to 0.06 S m$^{-1}$. In the band in which the SARAS 3 antenna operates, the real part of the relative complex permittivity is about 80 at 20°C. For water with conductivity less than about 0.3 S m$^{-1}$, the imaginary part is small compared to the real part. The magnetic permeability in water can be taken to be the same as in vacuum.

If the cone-disc antenna were on dry ground, with relative permittivity $\varepsilon_r$ = 5 and low conductivity of $\sigma$ = 0.002 S m$^{-1}$ the beam peaks at an elevation angle close to 33°. On fresh water, with relative permittivity $\varepsilon_r$ = 80 and low conductivity of $\sigma$ = 0.002 S m$^{-1}$, it drops to about 23°. On sea water, with substantially greater conductivity of $\sigma$ = 5 S m$^{-1}$, it comes down to about 11°. Compared to these, monopoles over perfect electrically conducting (PEC) ground will have their beams peaking towards the horizon. In the 40–90 MHz band the sky brightness is of order $10^3$ K and receiver noise is of order $10^2$ K. The measurement of sky spectral brightness using an antenna floating on the surface of a water body will have errors because the sky signal arrives at the antenna in multiple paths: a direct path and another reflected off the bottom of the water body. Additional source of error is multi-path interference between the receiver noise and its component that emerges from the antenna, reflects off the bottom of the water body, and re-enters the signal path. The magnitude of the spectral error is quantified by the attenuation in the E-field propagating 2-ways: to the bottom of the water body and back. The propagation constant (depth inside a medium at which the intensity of an incident wave falls to $1/e$ or 37% of its value) for EM waves in the frequency range 40–90 MHz, versus conductivity, is shown in Figure 11. In sea water, the propagation constant is small and in the range 2.5–3.6 cm in the 40–90 MHz band. For water with conductivity of 0.002 S m$^{-1}$, the propagation constant varies from 12 m at 90 MHz to 20 m at 40 MHz.

Using the propagation constant, we may compute the E-field attenuation for 2-way propagation for any depth of water. This is shown in Figure 12. EM waves incident from air into water or emerging from water into air would also suffer attenuations at the air-water interface owing to the impedance mismatch. For the low conductivities appropriate for most fresh water lakes, with conductivity less than 0.1 S m$^{-1}$, this attenuation is by factor about 4.4 dB for 2-way propagation of the E field. For air-sea water interface, the corresponding attenuation is somewhat greater than 10 dB.

As discussed in Section VI below, we have measured the S11 scattering matrix element for the antenna when it was floated on its raft on a lake. In water of depth 7 m, conductivity 0.007 S m$^{-1}$, S11 measurements showed a sinusoidal structure at level −48 dB below sky brightness temperature, with characteristic pattern expected for reflection at this depth. For this conductivity and depth, the attenuation in 2-way propagation is 15 dB, implying that in an $S$11 measurement only 33 dB of the E-field emerging from the antenna propagates down and could potentially be back reflected.

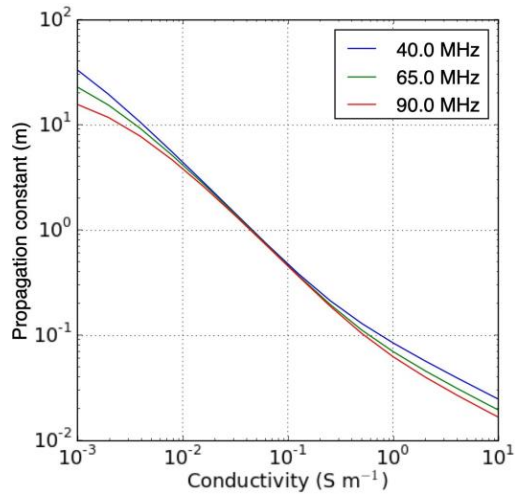

Fig. 11. Propagation constant for EM waves in water.

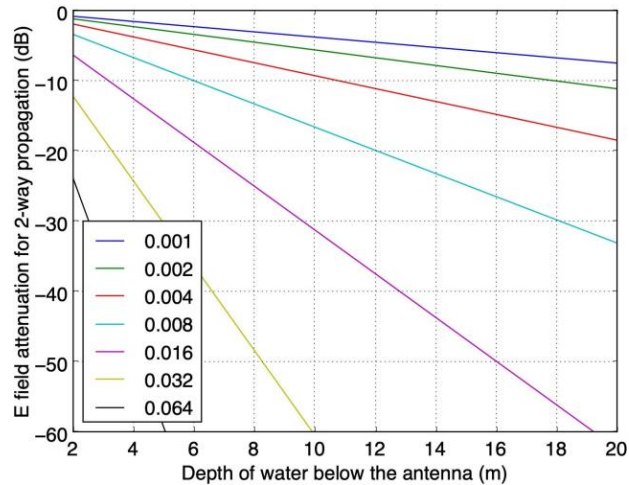

Fig. 12. The attenuation of the E field for 2-way propagation in water versus depth. The legend shows the values of conductivity, in S m$^{-1}$, for which the attenuations have been computed. It may be noted here that EM field strength would be attenuated by twice these values.

If their reflections from the bottom of the water body is to result in spectral structure less than a few mK, it is necessary that the total E-field attenuation needs to be about 60 dB. Thus a factor 27 dB of attenuation is needed in the 2-way propagation. In sea water even a depth of a meter provides required attenuation. In inland water bodies with conductivity of 0.02 S m$^{-1}$, the analysis shows that 6 m depth provides sufficient opacity; however, in fresh water lakes with significantly lower conductivity of 0.002 S m$^{-1}$, substantially greater depth of 40 m is required to achieve spurious-free measurement of Cosmic Dawn signals.

The velocity of EM waves in water depends primarily on the permittivity, but also the conductivity, and is substantially lower compared to that in free space. Therefore, the multi-path interference that arises in the case where one propagation path is in water will correspond to significantly large delays. The resulting multi-path interference between direct and reflected waves will manifest as spectral ripples in the antenna reflection efficiency, with periods given in Figure 13. It may be noted here that for depths beyond about 16 m, the spectral ripples will have periods lower than 1 MHz, and may easily be marginalized in the analysis since the Cosmic Dawn signals (see Figure 1) have substantially wider spectral structure. Thus, even for deployments of the SARAS 3 antenna in lakes with low conductivities, a depth of 16 m is adequate.

A potential issue with deployment on a water body is the impact of wind and waves. While translations and rotations of the antenna do not change the received signal, tilts of the antenna result in a change in the beam on the sky. The cosmic dawn signal is the same in all directions and, therefore, is received unchanged with beam tilts. However, the sky foreground component changes, but since that has a relatively smooth spectrum in all directions it is always subtracted out in the modeling of the foreground as a smooth function. Thus wind and waves do not limit the ability of the antenna to detect the cosmic dawn signal.

## V. FABRICATION OF THE SARAS 3 CONE-DISC ANTENNA

Some of the considerations that went into the fabrication were that the antenna needs to be (i) lightweight and portable for ease of transportation, (ii) structurally rigid for consistent and reliable performance, (iii) easy to assemble, and (iv) with joint between structural members in the direction of current flow. The surface of the cone and upper surface of reflector were made free from metallic projections due to screws and studs since they are observed to affect the antenna performance. Supporting structures that are not part of the EM design were made of electromagnetically transparent Styrofoam, and these were affixed using small quantities of metal-free glue.

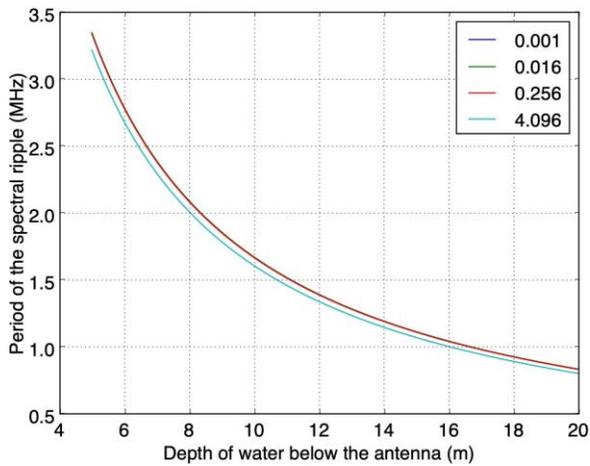

Fig. 13. The period the spectral ripple that will appear in measurement data with the SARAS 3 antenna on water, versus the depth of water.

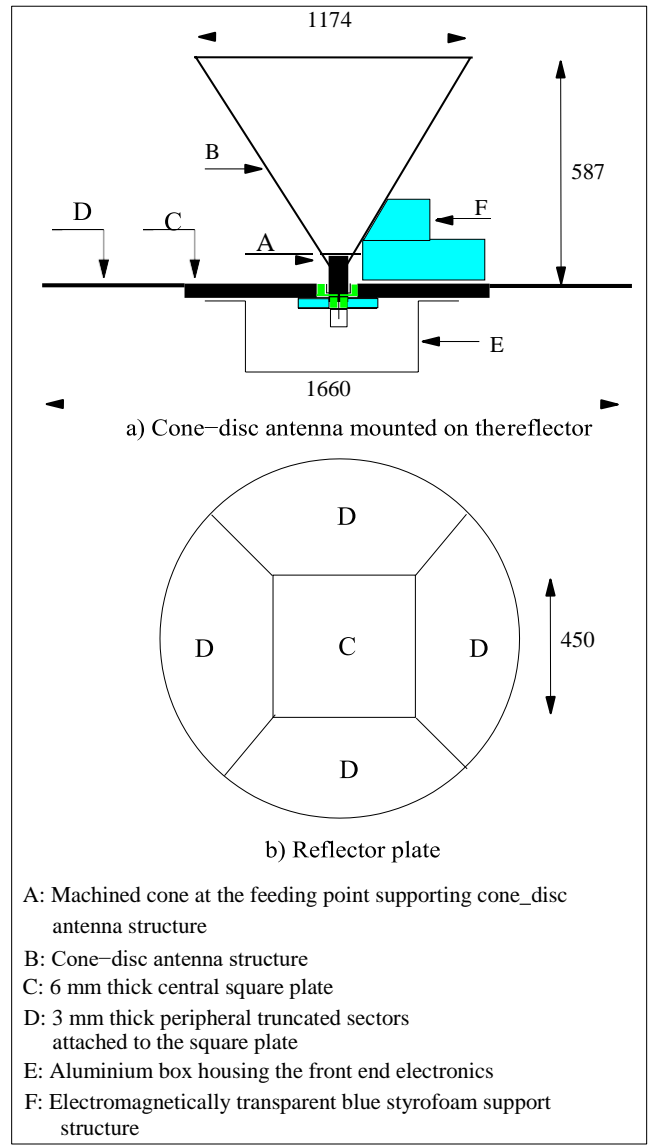

Fig. 14. Sketch of the structural components of the cone-disc antenna. All indicated dimensions are in mm.

Figure 14 shows the schematic of the antenna structure. It has two parts: (i) the conical monopole and (ii) a circular flat reflector. The inverted conical monopole is made of a 1.5mm thick sheet metal part and a machined block at the apex. The sheet metal is made of four discrete pieces reinforced using laser-cut-rings and vertical members as shown in Figure 15. The cone is of radius 587 mm at the top and 100 mm at the bottom. The inverted sheet-metal cone is sealed at the top with a circular aluminum disc: closing the structure showed an improvement in the performance of the antenna. A machined cone block (Figure 16) that forms the apex of the cone culminates at its bottom in a pin that fits into the jack of a UHF-SMA adaptor fitted at the center of the reflector plate. The pin connects the monopole to the receiver electronics. A step is provided at the top of the cone block for mounting the sheet metal part of the conical monopole. All joints in the conical monopole, including the rim of the top cover, are covered with aluminum tape to minimize discontinuity for the surface current flow and leakage.

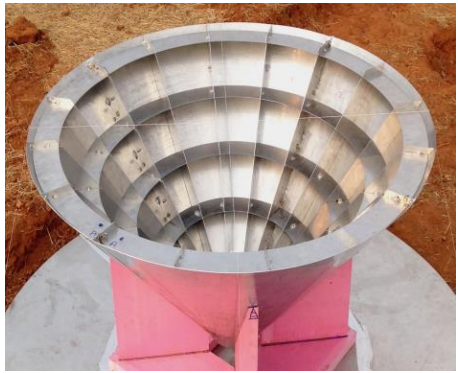

Fig. 15. Photograph showing the reinforcing members within the cone.

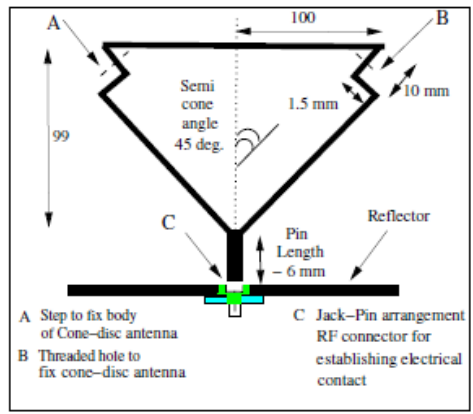

Fig. 16. Sketch of the machined cone block fitted to the apex of the conical monopole, providing electrical connection between the monopole and the receiver beneath.

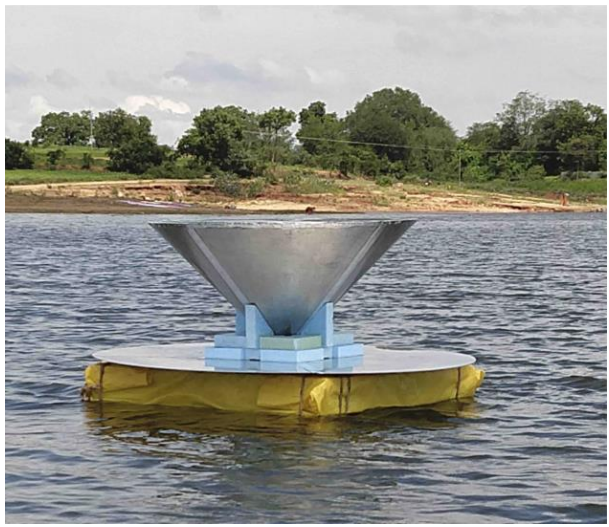

Fig. 17. Photograph of the fabricated antenna and its raft, deployed on the surface of a lake.

The reflector is a 3 mm thick circular plate of 0.83 m radius made of five discrete pieces for ease of assembly and transportation; the layout is shown in panel (b) of Figure 14. The antenna is placed on a 1.2 m square raft on water. The raft is made from standard blocks of Styrofoam, 1.250 m 0.6 m in size and 5 cm thick. These are electromagnetically transparent at the frequencies of operation, lightweight and has minimal water absorption. The front end receiver is housed at the center of the raft. The fabricated antenna, with the monopole element supported on the reflector plate by its styrofoam undergird, placed on its raft and deployed on a lake, is shown in Figure 17.

## VI. FIELD MEASUREMENTS OF THE RETURN LOSS AND EFFICIENCY

The reflection efficiency of the antenna may be estimated from an accurate measurement of its complex reflection coefficient. Usually, field measurement of $S11$ of an antenna is done by connecting the antenna to a 1-port vector network analyzer (VNA) using a long 50 Ω cable, which is calibrated either using standard precision loads or an electronic calibration unit (ECAL). This method was found to have insufficient accuracy, considering the science goal, for two reasons: (i) temporal drifts in cable characteristics between times of calibration and measurement, and (ii) parasitic effect of the cable on the antenna characteristics. For these two reasons, a precision $S11$ measurement technique with high cadence switching between calibrators and antenna has been used, as described below.

A N9912A Fieldfox Handheld RF analyzer from Keysight was placed within the receiver box kept below the antenna disc. A 4-way mechanical RF switch is mounted immediately below the antenna terminal and its common port is connected to the measurement port of the VNA using a very short cable. The antenna terminal is connected directly to one port of the switch; the other three ports of the 4-way switch are connected to precision terminations. Since the VNA is now below the antenna and floating on water well away from shore, ethernet- over-fiber is used to control it from a laptop on shore. The 4-way switch position is cycled through the calibration terminations and the antenna, and trace data acquired. The antenna data are calibrated using the termination data [24]. With this technique we have developed, we qualify $S11$ of goal: field measurements yield smooth $S11$ with residual rms the SARAS 3 antenna to be sufficiently smooth for the science of 7 ppm at spectral resolution of 0.7 MHz. The computed reflection efficiency, from the measurement data, is shown in Figure 18 along with the expectations from WIPL-D and FEKO.

The total efficiency of the antenna was estimated from spectral data recorded with the SARAS 3 radiometer, connected to the antenna, as the sky drifts overhead. The measured differential sky brightness yields the antenna efficiency; the formalism for this technique is described in [26]. The measurement data, after being corrected for the multiplicative bandpass of the receiver, are regressed against expected brightness predicted by GMOSS [27]. At each frequency, the measurements versus

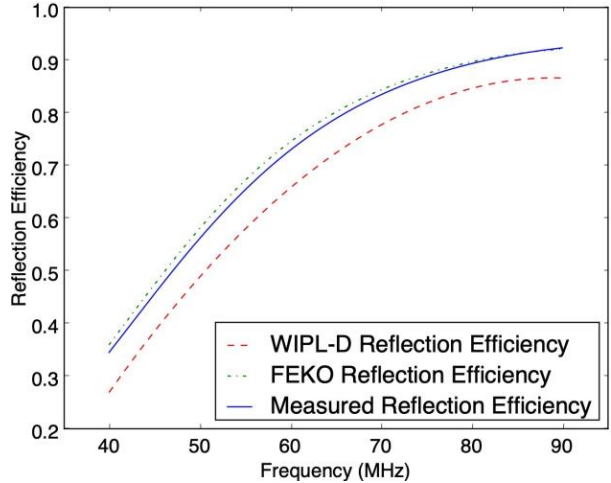

Fig. 18. Measured reflection efficiency for the optimized antenna design. Also shown for reference is the reflection efficiencies expected from WIPL-D and FEKO modeling.

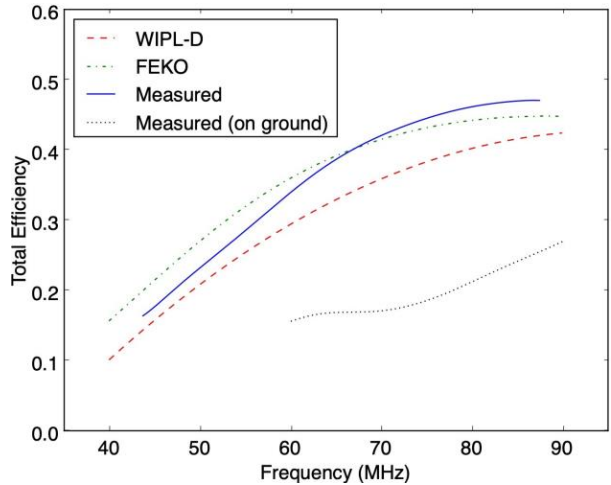

Fig. 19. Measured total efficiency for the SARAS 3 antenna built with the optimized cone-disc design, when afloat on a raft on a lake. Also shown for reference are the total efficiencies for this antenna on water expected from WIPL-D and FEKO modeling. The measured total efficiency of the antenna on dry ground is also shown.

GMOSS predictions, as a function of time, were fit with a straight line model. The slope of the linear fit provided the measure of the total efficiency of the antenna at that frequency, whereas the intercept measured the additive contributions in the measurements owing to emission from the water, etc. The total efficiency estimate is computed for each frequency independently. In Figure 19 we show the measured total efficiency compared with the expectations from the EM modeling. The total efficiency varies smoothly from about 10% at 40 MHz to about 45% at 90 MHz. As predicted, the total efficiency achieved with water as the medium beneath the antenna is observed to be substantially more than that when the antenna is placed on dry ground.

It may be noted that the sky brightness temperature is even if the efficiency is as low as 10% at the lowest frequency substantially more than the receiver noise temperature and so in the band, the signal to noise with which the cosmic dawn signal is received will not be diminished as a consequence of the low total efficiency. It is the signal-to-noise rather than signal amplitude that decides detection capability of the instrument.

The key advantage of placing the antenna on water rather than on ground is that (a) the improvement in efficiency results in improved signal to noise, and (b) the water provides a uniform medium that would only add an additive signal with a smooth spectrum, which would not confuse detection. Ground is potentially layered and inhomogeneous, resulting in emission with a complex spectrum that could potentially confuse any cosmic dawn signal. The complex shape of the efficiency on ground, shown in Figure 20, suggests exactly this.

## VII. Summary

We have successfully designed, constructed and field tested a floating cone-disc antenna with an octave bandwidth in the frequency range 43.75–87.5 MHz for the detection of the cosmological Cosmic Dawn signal, which is theoretically expected to appear in this band. The antenna characteristics have been cross-verified in two different electromagnetic tools (WIPL-D and FEKO) to ensure that it is not constrained by the limitations of individual softwares.

The innovations in the design solution involved (i) making the dimensions of the antenna less than half wavelength at all frequencies of the operating band, (ii) adopting a monopole design so that its feed point is at the antenna base and does not require a balun, (iii) making the structural dimensions vary uniformly from the feeding point in order to minimize surface current reflections at structural discontinuities, and (iv) carefully choosing the design frequency and structural parameters so as to keep any resonance beyond the band of operation. Most important, we designed the antenna for operation over the surface of a water body, for its high permittivity, instead of on ground, thus improving efficiency and also avoiding measurement errors which are frequency dependent from inhomogeneities in the medium beneath the antenna. The performance limitations from salinity and depth have been studied. The design naturally provides an achromatic beam.

Pin and jack arrangement of a typical RF connector has been adopted in our design for establishing electrical contact between the antenna and the receiver electronics. This has resulted in easy assembly and dismantling of the monopole element from the disc element during installation for remote field deployment. The antenna was placed on a raft made of standard blocks of electromagnetically transparent 5 cm thick blue Styrofoam.

The measurements of antenna performance were carried out while floating the antenna on water on its purpose-built raft, and ensuring an electromagnetic environment similar to that which will be used for observations of the celestial signal. Measurement of reflection efficiency was carried out by mounting the measuring instrument along with precision calibration standards immediately below the antenna disc in a receiver box, and switching between them and the antenna terminals at a rate sufficient to obtain the required accuracy and stability. The instrument control and data acquisition was carried out from shore with communications on an optical fiber. The accuracy of the $S11$ measurement was sufficient to validate this key antenna characteristic for the science goal. The total efficiency of the antenna was estimated by measuring the differential sky spectral brightness with the SARAS 3 radiometer, when it was connected to the antenna and observing the radio sky drift overhead. The total efficiency was measured to range from about 15% to 45% in the 43.75– 87.5 MHz octave band. The total efficiency achieved with water as the medium beneath the antenna is observed to be significantly more than that for dry ground. The antenna thus designed and constructed is potentially capable of providing a high-fidelity representation of the faint cosmological radio signal, along with foreground sky, to the receiver system.

## Acknowledgment

The authors thank the members of Electronics Engineering Group of the Raman Research Institute for assistance in fabrication and assembly of test and measuring equipment; in particular, S Kasturi who did the entire wiring and integration of the test and measurement equipment built inhouse. The Mechanical Engineering Services of the Institute led by Mohammed Ibrahim fabricated the antenna along with it Styrofoam support structure and raft, and assisted with deployments on lakes. Ashwathappa and members of the Gauribidanur Observatory helped with field measurements.